\documentclass[aps,prb,twocolumn,superscriptaddress,showpacs,byrevtex]{revtex4}
\usepackage{graphicx}
\begin{document}
\title{Split-off dimer defects on the Si(001)2~$\times$~1 surface}
\author{S.~R.~Schofield}
\email{steven@phys.unsw.edu.au}
\affiliation{Centre for Quantum Computer Technology, School of Physics, University of New South Wales, Sydney 2052, Australia}
\author{N.~A.~Marks}
\affiliation{Department of Applied Physics, University of Sydney, Sydney 2006, Australia}
\author{N.~J.~Curson}
\affiliation{Centre for Quantum Computer Technology, School of Physics, University of New South Wales, Sydney 2052, Australia}
\author{J.~L.~O'Brien}
\altaffiliation{Present address: Centre for Quantum Computer Technology, Department of Physics, University of Queensland, Brisbane 4072, Australia}
\affiliation{Centre for Quantum Computer Technology, School of Physics, University of New South Wales, Sydney 2052, Australia}
\author{G.~W.~Brown}
\affiliation{Los Alamos National Laboratory, Los Alamos, NM 87545, USA}
\author{M.~Y.~Simmons}
\affiliation{Centre for Quantum Computer Technology, School of Physics, University of New South Wales, Sydney 2052, Australia}
\author{R.~G.~Clark}
\affiliation{Centre for Quantum Computer Technology, School of Physics, University of New South Wales, Sydney 2052, Australia}
\author{M.~E.~Hawley}
\affiliation{Los Alamos National Laboratory, Los Alamos, NM 87545, USA}
\author{H.~F.~Wilson}
\affiliation{Department of Applied Physics, University of Sydney, Sydney 2006, Australia}
\date{May 5, 2003}
\begin{abstract}
Dimer vacancy (DV) defect complexes in the Si(001)$2\times1$ surface were investigated using high-resolution scanning tunneling microscopy and first principles calculations.  We find that under low bias filled-state tunneling conditions, isolated `split-off' dimers in these defect complexes are imaged as pairs of protrusions while the surrounding Si surface dimers appear as the usual ``bean-shaped'' protrusions.  We attribute this to the formation of $\pi$-bonds between the two atoms of the split-off dimer and second layer atoms, and present charge density plots to support this assignment.  We observe a local brightness enhancement due to strain for different DV complexes and provide the first experimental confirmation of an earlier prediction that the 1+2-DV induces less surface strain than other DV complexes.  Finally, we present a previously unreported triangular shaped split-off dimer defect complex that exists at S$\rm_B$-type step edges, and propose a structure for this defect involving a bound Si monomer.
\end{abstract}
\maketitle

\section{Introduction}
There are currently several exciting proposals to use the (001) surface of silicon for the construction of atomic-scale electronic devices, including single electron transistors~\cite{tu-ijcta-00-553}, ultra-dense memories~\cite{qu-n-01-265} and quantum computers~\cite{ka-na-98-133,ob-prb-01-161401}.  However, since any random charge or spin defects in the vicinity of these devices could potentially destroy their operation, a thorough understanding of the nature of crystalline defects on this surface is essential.  The Si(001) surface was first observed in real space at atomic resolution using scanning tunneling microscopy (STM) by Tromp \textit{et.~al.}\cite{tr-prl-85-1303} in 1985.  In this study they observed the surface consisted of rows of ``bean-shaped'' protrusions which were interpreted as tunneling from the $\pi$-bonds of surface Si dimers, thereby establishing the dimer model as the correct model for this surface. Since then, STM has been instrumental in further elucidating the characteristics of this surface, and in particular atomic-scale defects present on the surface\cite{ha-jvsta-89-2854,zh-srl-96-1449,ha-ss-00-156}. 

\begin{figure}
\centering
\includegraphics{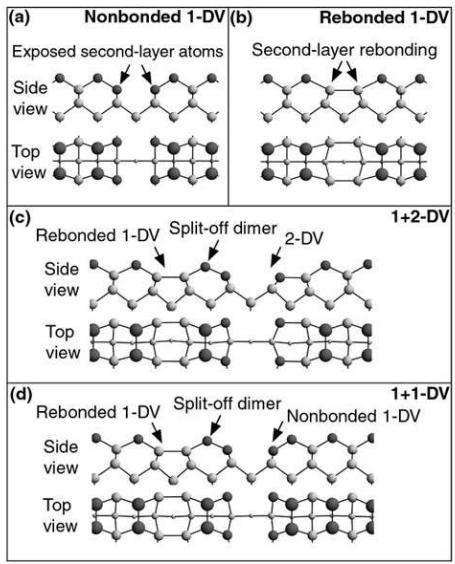}
\caption{Ball and stick models of dimer vacancy defects: (a) non-bonded 1-DV, (b) rebonded 1-DV, (c) 1+2-DV, and (d) 1+1-DV.  Si atoms that have a dangling bond are shaded black.  Height is indicated in the top views by the diameter of the balls, with the surface atoms having the largest diameter. The true minimum energy configurations for these structures involve buckling of the dimers in alternating directions along the dimer row.  However, since the dimers switch between their two possible buckling orientations at room temperature, the atomic positions shown here represent the average positions of the atoms.}
\label{def1}
\end{figure}

The simplest defect of the Si(001) surface is the single dimer vacancy defect (1-DV), shown schematically in Figs.~\ref{def1}(a) and \ref{def1}(b). This defect consists of the absence of a single dimer from the surface and can either expose four second-layer atoms (Fig.~\ref{def1}(a)) or form a more stable structure where rebonding of the second-layer atoms occurs~\cite{wa-prb-93-10497} as shown in Fig.~\ref{def1}(b).  While the rebonded 1-DV strains the bonds of its neighboring dimers it also results in a lowering of the number of surface dangling bonds and has been found to be more stable than the nonbonded structure.~\cite{ow-ss-95-L1042,wa-prb-93-10497}  Single dimer vacancy defects can also cluster to form larger defects such as the double dimer vacancy defect (2-DV) and the triple dimer vacancy defect (3-DV).  More complex clusters also form, the most commonly observed\cite{ko-prb-95-17269,wa-prb-93-10497} example is the 1+2-DV consisting of a 1-DV and a 2-DV separated by a single surface dimer, the so-called ``split-off dimer''.  The accepted structure of the 1+2-DV, as proposed by Wang \textit{et.~al.} based on total energy calculations,\cite{wa-prb-93-10497} is shown in Fig.~\ref{def1}(c) and consists of a rebonded 1-DV (left), a split-off dimer, and a 2-DV with a rebonding atom (right).  Recently we have observed another DV complex that contains a split-off dimer, called the 1+1-DV, which consists of a rebonded 1-DV and a nonbonded 1-DV separated by a split-off dimer, as shown in Fig.~\ref{def1}(d). 

Here we present a detailed investigation of DV defect complexes that contain split-off dimers.  Using high-resolution, low-bias STM we observe that split-off dimers appear as well-resolved pairs of protrusions under imaging conditions where normal Si dimers appear as single ``bean-shaped'' protrusions.  We show that this difference arises from an absence of the expected $\pi$-bonding between the two atoms of the split-off dimer but instead the formation of $\pi$-bonds between the split-off dimer atoms and second layer atoms. Electron charge density plots obtained using first principles calculations support this interpretation. We observe an intensity enhancement surrounding some split-off dimer defect complexes in our STM images and thereby discuss the local strain induced in the formation of these defects.  Finally, we present a model for a previously unreported triangular-shaped split-off dimer defect complex that exists at S$\rm_B$-type step edges.

\section{High-resolution variable-bias STM imaging of defect complexes}

Experiments were performed in two separate but identical variable temperature STM systems (Omicron VT-STM).  The base pressure of the ultra-high vacuum (UHV) chamber was $< 5\times10^{-11}$ mbar.  Phosphorus doped $10^{15}$ and $10^{19}$ cm$^{-3}$ wafers, orientated towards the [001] direction were used.  These wafers were cleaved into $2\times10$ mm$^2$ sized samples, mounted in sample holders, and then transferred into the UHV chamber.  Wafers and samples were handled using ceramic tweezers and mounted in tantalum/molybdenum/ceramic sample holders to avoid contamination from metals such as Ni and W.  Sample preparation\cite{sw-jvsta-89-2901} was performed in vacuum without prior {\em ex-situ} treatment by outgassing overnight at 850~K using a resistive heater element, followed by flashing to 1400~K by passing a direct current through the sample.  After flashing, the samples were cooled slowly ($\sim3$~K/s) from 1150~K to room temperature.  

\subsection{Split-off dimers}

The sample preparation procedure outlined above routinely produced samples with very low surface defect densities.  However, the density of defects, including split-off dimer defects, was found to increase over time with repeated sample preparation and STM imaging, as reported previously.\cite{ha-jvsta-00-1933}  It is known that split-off dimer defects are induced on the Si(001) surface by the presence of metal contamination such as Ni,~\cite{za-prl-95-3890} and W~\cite{ma-jjap-00-4518}.  The appearance of these defects in our samples therefore points to a build up of metal contamination, either Ni from in-vacuum stainless steel parts, or more likely W contamination from the STM tip.  After using an old W STM tip to scratch a $\sim$~1~mm line on a Si(001) sample in vacuum and then reflashing, the concentration of split-off dimer defects on the surface was found to have dramatically increased, confirming the STM tip as the source of the metal contamination.

\begin{figure}
\centering
\includegraphics{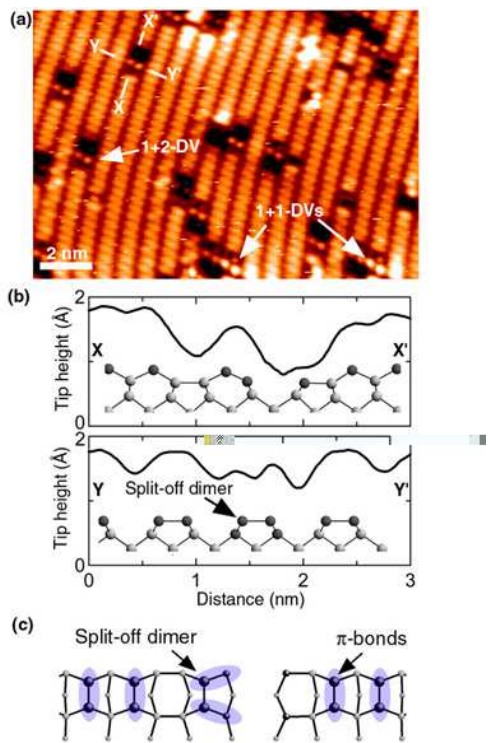}
\caption{A low bias filled-state STM image of a Si(001)2$\times$1 surface with split-off dimer defects is shown in (a).  Tunneling conditions for this image were $-1$~V sample bias and 0.8~nA tunnel current.  Line profiles are taken across a single 1+2-DV both parallel, X -- X$'$ (b), and perpendicular, Y -- Y$'$ (c), to the dimer row direction, as indicated in (a). The schematic (d) is a top view ball and stick model of a 1+2-DV with the approximate positions of $\pi$-bonds indicated by shaded ellipses.}
\label{SODs}
\end{figure}

Figure~\ref{SODs} shows an STM image of a Si(001) surface containing a $\sim$~10\% coverage of split-off dimer defects.  The majority of the defects in this image can be identified as 1+2-DVs, however, two 1+1-DVs are also present, as indicated.   The most striking feature of this image is the difference in appearance of the split-off dimers in contrast to the surrounding normal surface dimers.  Each split-off dimer in this image appears as a double-lobed protrusion, while the surrounding normal Si dimers each appear as a single ``bean-shaped'' protrusion, as expected at this tunneling bias.~\cite{ha-prb-99-8164}  Line profiles taken across a 1+2-DV both parallel and perpendicular to the dimer row direction are shown in Fig.~\ref{SODs}(b).  The line profile parallel to the dimer row direction agrees with previously reported profiles over 1+2-DVs and fits well with the accepted structure,~\cite{zh-srl-96-1449,ow-ss-95-L1042} as shown by the overlayed ball and stick model.  The line profile taken perpendicular to the dimer row direction, however, clearly shows that the split-off dimer of this defect is separated into two protrusions while the neighboring Si dimers are single protrusions.  This is the first recognition and explanation of split-off dimers appearing as double-lobed protrusions.

To understand why split-off dimers appear as double-lobed protrusions we must consider the structure of these defects shown in Figs.~\ref{def1}(c) and \ref{def1}(d).  Normally Si(001) surface dimers appear as ``bean-shaped'' protrusions in STM images because the dangling bonds of each Si dimer atom mix to form a $\pi$-bond between the two dimer atoms.  However, if we examine the split-off dimer structure closely (Figs.~\ref{def1}(c) and \ref{def1}(d)) we see that unlike normal surface dimers, the split-off dimer has two nearest neighbor second layer atoms that each have a dangling bond.  The separation distance between the split-off dimer atoms and these second layer atoms is sufficiently close to allow the formation of $\pi$-bonds.  The resulting four-atom structure can therefore be referred to as a \textit{tetramer}. We propose that the four dangling bonds of the split-off dimer tetramer interact primarily along the backbonds between the split-off dimer atoms and the second layer atoms to form $\pi$-bonds down the backbonds, as drawn schematically in Fig.~\ref{SODs}(c).  These two spatially separated $\pi$-bonds therefore lead to the double-lobed appearance of the split-off dimers under low bias filled-state tunneling conditions, which we confirm in section~\ref{theory1} with charge density calculations.

\begin{figure}
\centering
\includegraphics{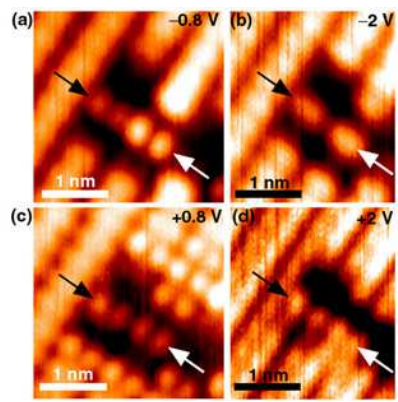}
\caption{Variable bias STM images of a 1+2-DV adjacent to a 1+1-DV.  The split-off dimer of the 1+2-DV is indicated with a black arrow, while the split-off dimer of the 1+1-DV is indicated by a white arrow.  All four images were acquired with 0.13~nA tunnel current and the sample bias for each image is (a) $-0.8$~V, (b) $-2$~V, (c) $+0.8$~V, (d) $+2$~V.}
\label{SODv}
\end{figure}

In an attempt to fully characterize the appearance of these split-off dimers in STM images, we have performed a series of experiments observing split-off dimers with changing STM sample bias.  Figure~\ref{SODv} summarizes our results, showing images where a 1+2-DV and a 1+1-DV located next to each other are observed at four different sample biases -- two filled-state images and two empty-state images.  In the filled-state image of Fig.~\ref{SODv}(a) we see that at $-0.8$~V the split-off dimer of both the 1+2-DV and the 1+1-DV appear as double-lobed protrusions similar to those in Fig.~\ref{SODs}(a).  However, when the filled-state bias is increased in magnitude to $-2$~V, Fig.~\ref{SODv}(b), the split-off dimers become single-protrusions and appear very similar to the surrounding normal Si surface dimers.  This is because as the bias magnitude is increased towards $-2$~V, the dimer $\sigma$-bond and bulk states contribute increasingly to the tunneling current~\cite{ha-prb-99-8164} and the image of the split-off dimer reverts to the bean-shaped protrusion in the same manner as normal surface Si dimers.  In both of the empty-state images, Figs.~\ref{SODv}(c) and \ref{SODv}(d), acquired at +0.8~V and +2~V, respectively, the appearance of the split-off dimers is very similar to that of the surrounding normal surface dimers.  This is because under empty-state tunneling conditions electrons tunnel into the $\pi^*$-antibonding orbitals of the dimers, resulting in the normal Si dimers appearing as double-lobed protrusions.~\cite{qi-prb-99-7293}  It is therefore only under low bias magnitude filled-state tunneling conditions that split-off dimers appear significantly different to the surrounding normal Si surface dimers.

\subsection{Experimental observation of surface strain in complex defects}
\label{strainsection}

\begin{figure}
\centering
\includegraphics{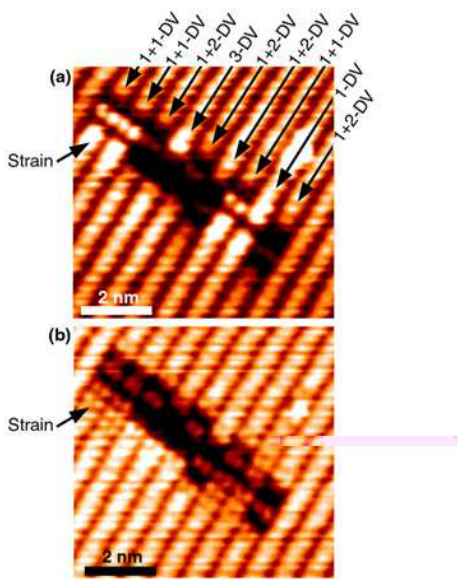}
\caption{Filled and empty-state STM images ($-1.2$~V, $+1.6$~V, 0.15~nA) of a short chain of DVs in a Si(001) surface.  The individual defects are (from top left to bottom right): 1+1-DV, 1+1-DV, 1+2-DV, 3-DV, 1+2-DV, 1+2-DV, 1+1-DV, 1-DV, and 1+2-DV.  Note the strain-induced brightening of the 1-DV and 1+1-DVs in the filled-state (a) and the corresponding darkening in the empty-state (b)}
\label{strain}
\end{figure}

Another noticeable feature of Figs.~\ref{SODs}(a) and \ref{SODv}(a) is the enhanced brightness of the 1+1-DV compared to the 1+2-DV.  This is a reproducible effect that we attribute to an increased amount of surface strain induced by the 1+1-DV.  Figure~\ref{strain} shows a series of adjacent defects forming a short vacancy line channel in the surface.  This channel is composed of individual 1-DV, 3-DV, 1+2-DV, and 1+1-DV defects (see figure caption).  In the filled-state image, Fig.~\ref{strain}(a), there is a clear brightening of the dimers on one end of the 1+1-DVs and the dimers on both ends of the 1-DV, which is not present for the 1+2-DVs.  In the empty-state image of the line of defect complexes, Fig.~\ref{strain}(b), we notice that there is a darkening of the same dimers that are enhanced in the filled-state image.

Owen \textit{et.~al.},~\cite{ow-ss-95-L1042} have shown using low bias STM and first principles calculations, that the dimers neighboring a rebonded 1-DV are enhanced in low bias filled-state STM images due to the strain induced by the defect shifting the surface states upwards in energy toward the Fermi energy.  This effect can be seen for the 1-DV in Fig.~\ref{strain}(a), where the neighboring dimers in the same row as the 1-DV are enhanced in intensity, with the magnitude of the enhancement decaying with distance from the 1-DV.  A very similar enhancement can be seen around the 1+1-DV sites in this image, with the split-off dimer in particular appearing much brighter than the surrounding normal surface dimers.  However, for the 1+1-DV only the dimers on one end of the defect are enhanced in intensity while the dimers on the other end of the defect are not.  This observation can be readily explained since the 1+1-DV is composed of a rebonded 1-DV adjacent to a nonbonded 1-DV (Fig.~\ref{def1}(d)) and Owen~\textit{et.~al.}~\cite{ow-ss-95-L1042} have shown that while the rebonded 1-DV results in strain-induced image enhancement, the nonbonded 1-DV does not.  The observation of an asymmetric strain-induced enhancement of the 1+1-DV in Fig.~\ref{strain}(a) can therefore be taken as an experimental confirmation of the structure of this defect (Fig.~\ref{def1}(d)) and the first application of the method of Owen~\textit{et. al.}~\cite{ow-ss-95-L1042} for identifying strain in more complex surface defect structures.

The fact that the 1+2-DV causes no enhancement of its neighboring dimers over the surrounding normal surface dimers suggests that the 1+2-DV, unlike the 1-DV and 1+1-DVs, does not increase the strain of the surface.  This at first seems strange, since the 1+2-DV involves a rebonded 1-DV similar to the 1+1-DV structure.  However, Wang \textit{et.~al.}~\cite{wa-prb-93-10497} have shown, using total energy calculations, that the junction formed between the 1-DV and the 2-DV to create the 1+2-DV releases the surface strain that is present when these two defects exist separately from one another.  The STM data that we have presented here is therefore the first experimental verification of this calculation.  The fact that both the 1-DV and the 1+1-DV show local enhancement due to strain, while the 1+2-DV does not, indicates that the 1+2-DV structure induces less local strain than the 1-DV.  

In their paper, Owen \textit{et.~al.} do not present empty state STM images, nor do they consider empty states in their tight binding calculations.  In Fig.~\ref{strain}(b), we show an empty state image of the same line of defects shown in Fig.~\ref{strain}(a).  Interestingly, in this empty state image the dimers that were enhanced in brightness surrounding the 1-DV and 1+1-DVs in the filled-state image are less bright than the surrounding Si dimers in the empty-state image.  This suggests that the strain associated with these defects causes the lowest unoccupied molecular orbital (LUMO) of the adjacent dimers to also shift higher in energy, away from the Fermi energy.  

\section{Density functional characterization of defect complexes}
\label{theory1}

To confirm the interpretation of our STM images, we have performed first principles electronic structure calculations of both the 1+2-DV and 1+1-DV complexes using the Car-Parrinello Molecular Dynamics program.~\cite{cpmd}  Valence electrons were described using Goedecker pseudopotentials~\cite{go-prb-96-1703} expanded in a basis set of plane waves with an energy cutoff of 18 Rydbergs and the exchange-correlation functional was of the BLYP form.~\cite{be-pra-88-3098,le-prb-98-785} Slab calculations contained between 124 and 128 Si atoms in a $31.070\times7.675\times19.253$~\AA$^3$ supercell, corresponding to six layers of vacuum in the $z$-direction, and all calculations were performed with gamma point sampling of the Brillouin zone only.  A reference calculation was performed with no surface vacancies and assuming the $p(2\times2)$ structure in which the dimers buckled alternately along the row.  A single 256 atom calculation with a duplication along the y-axis confirmed that the effect of dispersion across the rows is minor as has been noted elsewhere.~\cite{po-jvstb-87-945}

Both zero temperature geometry optimization and high temperature molecular dynamics calculations were used to explore a variety of surface and 
second-layer bonding configurations for the 1+2-DV and 1+1-DV.  The results confirm the
configurations in Figs.~\ref{def1}(c) and \ref{def1}(d) are the lowest energy geometries of both defect complexes.  The dimers are drawn symmetric in these schematics, however, the true minimum energy structure at zero temperature involved charge-transfer buckling of the Si dimers.  It is well known that at room 
temperature the barrier is sufficiently small for 
the dimers to flip-flop between the two equivalent 
configurations.~\cite{ra-prb-95-14504,wo-prl-92-2636}
Our calculations show that the split-off dimer tetramer also has
two symmetrically equivalent buckling configurations, 
with charge transfer between the atoms of the tetramer 
buckling adjacent atoms in alternate directions.
By analogy with the normal dimers we can expect 
room-temperature STM measurements of the tetramer
to image the average of the two configurations.
The chemical potential was determined from a 
512-atom bulk calculation, which yielded a
formation energy of 0.85~eV for the 1+2-DV, similar 
to the value of 0.65~eV computed by Wang \textit{at. al}.~\cite{wa-prb-93-10497}
The 1+1-DV formation energy has not been previously reported,
and we found it to be 1.13~eV.  We note that this value is high, but this is consistent with the rarity of observation of the 1+1-DV in STM
experiments.

\begin{figure}
\includegraphics{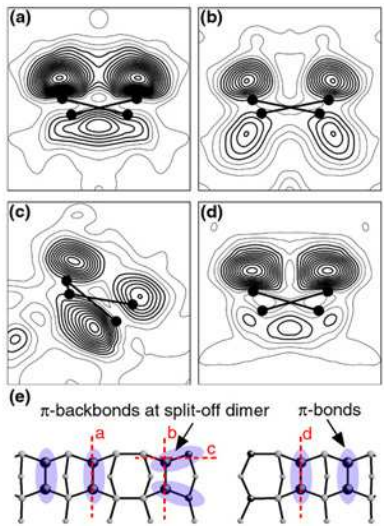}
\caption{Cross-section electron density plots for filled states within 0.25~eV of 
the highest occupied molecular orbital (HOMO) for several cuts through the 
1+2-DV complex. The planes a,b,c and d through the top-view ball and 
stick model (e) indicate the direction and position of the cuts, and the 
shaded ellipses indicate the $\pi$-bonding as inferred from the electron 
density (see text). Each electron density plot is an average of both 
buckling configurations, and the atomic positions and bonds are shown as 
black balls and sticks. The slices are (a) rebonded 1-DV edge dimer, (b) 
split-off dimer, (c) split-off dimer backbonds, (d) 2-DV edge dimer.}
\label{1+2-dv}
\end{figure}

In Fig.~\ref{1+2-dv} we present a series of calculated electron density slices 
through various regions of the 1+2-DV marked by (a), (b), (c), and (d) in the ball and stick schematic. The charge density shown in the 
figure is the sum of the occupied Kohn-Sham orbitals within 0.25~eV of 
the highest occupied molecular orbital (HOMO). Taking into account the 
$\sim0.5$~eV surface band gap of Si(001) and the n-type doping of the 
experimental samples, these states correspond approximately to the 
accessible states for a $\sim0.75$~V sample bias and can therefore be 
directly compared to the experimental data in Fig.~\ref{SODv}(a), which was 
acquired with a $-0.8$~V sample bias.

The four charge density slices in Fig.~\ref{1+2-dv} show: Fig.~\ref{1+2-dv}(a) the 1-DV edge dimer, 
Fig.~\ref{1+2-dv}(b) the split-off dimer, Fig.~\ref{1+2-dv}(c) the backbond of the split-off dimer, and 
Fig.~\ref{1+2-dv}(d) the 2-DV edge dimer, as indicated schematically in Fig.~\ref{1+2-dv}(e). The charge 
densities of both buckling configurations of the dimers and backbond atoms are averaged, 
and the positions of the dimer and tetramer atoms are shown superimposed 
in both buckling configurations. In the case of the backbonds, the two 
configurations are not coincident, and so the atoms and bonds are shown 
in projection onto the plane in Fig.~\ref{1+2-dv}(c). The 1-DV edge 
dimer in Fig.~\ref{1+2-dv}(a) shows a clear three-lobed character with 
significant overlap between the up-atom charge density of the two 
buckling orientations, and a single lobe beneath the plane of the 
surface at the mid-point of the dimer. Density functional calculations 
by Hata~\textit{et.~al.}~\cite{ha-prb-99-8164} and tight-binding Green's function calculations by 
Pollman~\textit{et.~al.}~\cite{po-jvstb-87-945} have separately identified this three-lobed feature as 
being characteristic of $\pi$-bonding in flip-flop dimers on the silicon 
surface, and we can therefore take this three-lobed feature as a signature of 
$\pi$-bonding in this work. The backbond of the split-off dimer in Fig.~\ref{1+2-dv}(c) connects a first-layer atom to a second-layer atom and also shows a 
three-lobed structure. By analogy with the surface dimer in Fig.~\ref{1+2-dv}(a) 
we characterize this bond as having $\pi$-character and have indicated this by the 
shaded ellipse (c) shown in Fig.~\ref{1+2-dv}(e). The split-off dimer itself in Fig.~\ref{1+2-dv}(b), however, does not exhibit three-lobed character. Instead, the 
split-off dimer has four lobes; two located above the up-atoms of the 
dimer in each buckling configuration, and a second pair of spatially 
separated lobes beneath the bond. The calculations thus show that 
$\pi$-bonding occurs down the backbonds of the split-off dimer, but not 
across the dimer itself.   The absence of the $\pi$-bond across the 
split-off dimer correlates with the double-protrusions observed in the 
STM images.  Finally, we also consider the charge density of the 2-DV edge 
dimer, Fig.~\ref{1+2-dv}(d), and note that it also exhibits three-lobed character, 
indicative of $\pi$-bonding.   This gives the 2-DV edge dimer a bean-shaped 
appearance in the STM image, as for the 1-DV dimer in Fig.~\ref{1+2-dv}(a).

\begin{figure}
\includegraphics{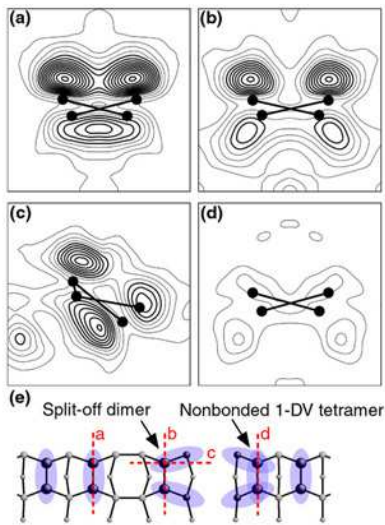}
\caption{Cross-section electron density plots for filled states within 0.25~eV of 
the HOMO for several cuts through the 
1+1-DV complex. The planes a,b,c and d through the top-view ball and 
stick model (e) indicate the direction and position of the cuts, and the 
shaded ellipses indicate the $\pi$-bonding as inferred from the electron 
density (see text). Each electron density plot is an average of both 
buckling configurations, and the atomic positions and bonds are shown as 
black balls and sticks. The slices are (a) rebonded 1-DV edge dimer, (b) 
split-off dimer, (c) split-off dimer backbonds, (d) nonbonded 1-DV edge 
dimer.}
\label{1+1-dv}
\end{figure}

A similar situation exists for the 1+1-DV charge density slices shown in 
Fig.~\ref{1+1-dv}. The first three charge density slices, Figs.~\ref{1+1-dv}(a) -- \ref{1+1-dv}(c), are analogous to 
the slices for the 1+2-DV As was the case for the 
1+2-DV, the rebonded 1-DV edge dimer, Fig.~\ref{1+1-dv}(a) and the split-off dimer 
backbonds, Fig.~\ref{1+1-dv}(c) exhibit three-lobed $\pi$-like character, while the split-off 
dimer, Fig.~\ref{1+1-dv}(b) exhibits four-lobed character, consistent with  an end-on view 
of $\pi$-bonding down the backbonds. Finally, another slice is presented in Fig.~\ref{1+1-dv}(d), which is through the 
nonbonded 1-DV edge dimer as indicated schematically in Fig.~\ref{1+1-dv}(e).  It can be seen that the nonbonded 1-DV edge dimer 
appears quite different to the charge density slices discussed so 
far. In particular, we notice that the nonbonded 1-DV edge dimer has a 
much reduced charge density compared to the other slices, Fig.~\ref{1+1-dv}(a) -- \ref{1+1-dv}(c).  
Examination of the structure identifies strain as the characteristic 
that differentiates the dimer in Fig.~\ref{1+1-dv}(d) from the other dimers.  Since the 
dimer in Fig.~\ref{1+1-dv}(d) is part of a tetramer, one might expect its appearance to 
resemble the split-off dimer which is also part of the tetramer shown 
Figs.~\ref{1+1-dv}(b) and \ref{1+1-dv}(c).  However, a detailed examination of the simulated structure 
reveals that the nonbonded 1-DV tetramer is relaxed, since there is one adjacent 
dimer present, while the split-off tetramer is highly strained 
because of the rebonding in the second-layer.  Since the nonbonded 1-DV 
tetramer is much less strained, its occupied states lie further from the 
Fermi level, explaining the charge reduction observed in calculations in 
Fig.~\ref{1+1-dv}(d).  As discussed in Ref.~\onlinecite{dab-ch3}, the minimum energy arrangement of the 
electrons in a tetramer is one where the $\pi$-states are delocalized across 
the four atoms, to form three bonding segments, as indicated by the 
ellipses in Fig.~\ref{1+1-dv}(e). The charge density slice of Fig.~\ref{1+1-dv}(d) is consistent 
with such an arrangement where the charge density is shared between 
$\pi$-like bonds on both backbonds and across the dimer atoms. We conclude 
that this charge density arrangement forms for the nonbonded 1-DV 
tetramer because it is allowed to relax.  In the case of the split-off dimer, the 
tetramer is constrained by the rebonding and instead forms a higher 
energy configuration in which the $\pi$-bonds conjugate to form two $\pi$-bonds 
down its backbonds.

\section{New step edge defect}

\begin{figure}
\includegraphics{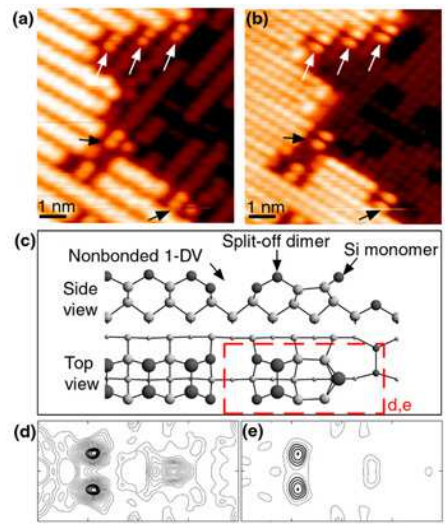}
\caption{(a), (b) Filled- and empty-state images ($\pm$1.2~V) of DV defects at an S$\rm_B$-type step edge.  White arrows indicate S$\rm_B$-DVs,~\cite{ko-prb-96-10308} while black arrows point to a previously unreported defect that exhibits a third protrusion in the filled-state giving it a triangular appearance.  We propose the structure (c) as a model for this defect.  Calculated charge density slices at a constant $z$-height for the dashed region of (c) are shown in (d) and (e) (for Kohn-Sham orbitals summed over 0.45~eV below the HOMO and 0.45~eV above the LUMO, respectively).  These contour slices are in good agreement with the STM images in (a) and (b), in particular predicting the correct spacing of 6.4~\AA\ between the split-off dimer and third protrusion and also the disappearance of the third protrusion in the empty-state.  The horizontal tic-marks in (d) and (e) indicate the dimer positions on the defect-free surface.}
\label{triangular}
\end{figure}

Having presented a detailed understanding of the electronic structure of previously observed split-off dimer defects in the Si(001) surface using both STM and first-principles calculations, we now turn our attention to elucidating the structure of a previously unreported split-off dimer defect.  
In Figs.~\ref{triangular}(a) and \ref{triangular}(b) we show filled- and empty-state STM images of DV defects at a single-layer S$\rm_B$-type step edge.  
At the top of these images white arrows indicate are three defects known as S$\rm_B$-DVs, which are rebonded 1-DVs at the step edge, which leave a single split-off dimer as the last dimer before the lower terrace begins.~\cite{ko-prb-96-10308}  As was the case for the 1+1-DV and 1+2-DV, the split-off dimers in S$\rm_B$-DVs appear as double-lobed protrusions under low-bias filled-state imaging conditions, Fig.~\ref{triangular}(a).  
At the bottom of Fig.~\ref{triangular}(a) two similar DV complexes can be observed, as indicated by black arrows, however these defects have a third protrusion giving them a triangular appearance.  In empty-state imaging, Fig.~\ref{triangular}(b), however, the additional third feature is not present.  These triangular-shaped defects have not been reported on the Si(001) surface before and most likely arise due to the presence of W contamination.

Our proposed structural model of the triangular-shaped defects in Fig.~\ref{triangular}(a) is shown in Fig.~\ref{triangular}(c).  This model consists of a nonbonded 1-DV defect at an S$\rm_B$-type step edge, followed by a rebonded split-off dimer and a bound Si monomer.  Swartzentruber has previously observed Si monomers on the Si(001) surface using high-resolution STM after depositing a few percent of a monolayer of Si atoms to the surface.~\cite{sw-jcg-98-1}  These monomers were bound at rebonded S$\rm_B$-type step edges, confirming the minimum energy binding position predicted by first principles calculations.  The binding position of the monomer in our proposed structure, Fig.~\ref{triangular}(c), is essentially the same position observed by Swartzentruber, with the difference being the presence of the DV defect adjacent to the step edge.  Swartzentruber also observed that the Si monomers bound at S$\rm_B$-type step edges were visible  in one bias polarity (empty-state) but invisible in the other (filled-state).  Our images reveal a similar effect, however the feature we observe appears in filled-state images while being invisible in empty-state images.

We have performed first-principles calculations to produce charge density contours for our proposed structure.  Figure 7(d) shows a constant $z$-height contour slice taken 1.2~\AA\ above the monomer for occupied Kohn-Sham orbitals within 0.45~eV of the HOMO.
We see in this charge density contour slice the two lobes expected for the split-off dimer as well as a third lobe due to the bound monomer.  Moreover, the distance between the split-off dimer lobes and the monomer lobe is 6.4~\AA\, in agreement with the separation seen in the STM image.  In Fig.~\ref{triangular}(e) we show an empty-state slice taken at the same $z$-height and summed over Kohn-Sham orbitals up to 0.45~eV above the LUMO.  In this contour the double lobe of the split-off dimer is still present but the monomer lobe is significantly lessened in intensity.  The results of our first-principles calculations therefore give good agreement between our proposed structure and the observed defect.  The presence of the split-off dimer must therefore be responsible for the reversal of the filled- and empty-state monomer characteristics when compared to those observed for monomers bound to rebonded S$\rm_B$-type step edges.

\section{Summary}
We have investigated split-off dimers on the Si(001)2$\times$1 surface using high resolution STM and first principles calculations.  We find that split-off dimers form $\pi$-bonds with second layer atoms which gives them a double-lobed appearance in low bias filled-state STM images.  We apply the method of Owen~\textit{et. al.}~\cite{ow-ss-95-L1042} for identifying local areas of increased surface strain to dimer vacancy defect complexes and thereby present the first experimental confirmation of the predicted strain relief offered by the 1+2-DV.  Finally, we have presented a previously unreported triangular-shaped defect on the Si(001) surface and a proposed model for this structure involving a bound Si monomer.

\end{document}